\documentstyle[12pt,epsf]{article}

\def\be{\begin{equation}}
\def\ee{\end{equation}}
\def\bea{\begin{eqnarray}}
\def\eea{\end{eqnarray}}

\begin{document}

\begin{flushright}
hep-th/0104210
\end{flushright}

\pagestyle{plain}

\def\e{{\rm e}}
\def\haf{{\frac{1}{2}}}
\def\tr{{\rm Tr\;}}
\def\goes{\rightarrow}
\def\ie{{\it i.e.}, }
\def\tcl{T_{\rm cl}}
\def\Goes {\Rightarrow}
\def\CX{{\cal X}}

\begin{center}
\vspace{2cm}
{\Large {\bf Electrodynamics On Matrix Space:}}
\vspace{.5cm}

{\Large {\bf Non-Abelian By Coordinates}}

\vspace{1cm}

Amir H. Fatollahi 
\footnote{
On leave from: Institute for Advanced Studies in Basic Sciences (IASBS), 
Zanjan, Iran.}

\vspace{.5cm}

{\it Dipartimento di Fisica, Universita di Roma ``Tor Vergata",}\\ {\it
INFN-Sezione di Roma II, Via della Ricerca Scientifica, 1,}\\ {\it 00133,
Roma, Italy}

\vspace{.3cm}

{\sl fatho@roma2.infn.it}

\vskip .5 cm
\end{center}

\begin{abstract}
We consider the dynamics of a charged particle in a space whose
coordinates are $N\times N$ hermitian matrices. Putting things in the
framework of D0-branes of String Theory, we mention that the
transformations of the matrix coordinates induce non-Abelian
transformations on the gauge potentials. The Lorentz equations of motion
for matrix coordinates are derived, and it is observed that the field
strengths also transform like their non-Abelian counterparts. The issue of
the map between theory on matrix space and ordinary non-Abelian gauge
theory is discussed. The phenomenological aspect of ``finite-N
non-commutativity" for the bound states of D0-branes appears to be very
attractive. 
\end{abstract}

\newpage

{\bf Electrodynamics On Matrix Space:} We begin with the dynamics of a
charged point particle in a space whose coordinates are $N\times N$
hermitian matrices, such as 
\bea
X^i=X^i_a T^a, \;\;\;i=1,\cdots, d,\;\;\;a=1,\cdots, N^2
\eea
in which $T^a$ are the basis for hermitian matrices (\ie the generators
of $U(N)$). The action may be in the form of 
\bea\label{SX} 
S[X]=\int dt\; \tr (\haf m \dot X_i \dot X^i + q \dot X^i A_i(X,t)  - q
A_t(X,t)-V(X))&
\eea
which can be obtained simply by replacing ordinary coordinates, $x$, by
their matrix form $X$, in the action $S[x]=\int dt (\haf m \dot x_i \dot
x^i - q \dot x^iA_i(x,t) - q A_t(x,t)-V(x))$, simply added by a ``Tr" 
on the matrix structure. Besides we assume that the gauge potentials
$(A_t(X,t),A_i(X,t))$ have functional dependence on the matrix
coordinates $X$, and to put things simple (and natural) the $\tr$ should
be calculated by ``symmetrization prescription" on the matrices $X$. By
symmetrization prescription we mean symmetrization on the all of $X$'s
appearing in the potentials; this can be obtained by the so-called
``non-Abelian Taylor expansion," as
\bea\label{NTE}
A_\mu(X,t)&=&
A_\mu(x,t)|_{x\goes X}\equiv\exp[X^i\partial_{x^i}] 
A_\mu(x,t)\nonumber\\
&=&\sum_{n=0}^\infty
\frac{1}{n!} X^{i_1}\cdots X^{i_n} 
(\partial_{x^{i_1}}\cdots\partial_{x^{i_n}})
A_\mu(x,t)|_{x=0},
\eea
with $\mu=t,i$. In the above expansion the symmetrization is recovered via
the symmetric property of the derivatives inside the term
$(\partial_{x^{i_1}}\cdots\partial_{x^{i_n}})$. Now we have an action with
enhanced degrees of freedom, from $d$ in ordinary space, to $d\times
N^2$ in space with matrix coordinates.

The fate of the $U(1)$ symmetry of the action $S[x]$, with 
transformations as
\bea
A_\mu(x,t)\goes A'_\mu(x,t)=A_\mu(x,t)- \partial_\mu \Lambda(x,t),
\eea
in the new action $S[X]$ is interesting. One can see that the
action $S[X]$ is also symmetric under similar transformations, as
\bea\label{UOT}
A_t(X,t)&\goes& A'_t(X,t)=A_t(X,t)-\partial_t\Lambda(X,t)\nonumber\\
A_i(X,t)&\goes& A'_i(X,t)=A_i(X,t)+\delta_i\Lambda(X,t),
\eea
in which $\delta_i$ is the functional derivative $\frac{\delta}{\delta
X^i}$. Consequently one obtains:
\bea\label{VSX}
\delta S[X] \sim q \int dt\; \tr \bigg(
\partial_t \Lambda(X,t) +
\dot X^i \delta_i\Lambda(X,t) \bigg) \sim q \int dt\; \tr \bigg(
\frac{d\Lambda(X,t)}{dt}\bigg) \sim 0.
\eea

{\bf D0-Brane Picture:} Since we are performing symmetrization in gauge
potentials $(A_t,A_i)$, the symmetric parts of the potential $V(X)$ can be
absorbed in a redefinition of $A_t(X,t)$. So the interesting parts of
$V(X)$ contain ``commutators" of coordinates, in an expansion could be
presented as
\bea
V(X)=\underbrace{[X^i,X^j]+X^i[X^j,X^k]}_{{\rm
traceless\;or\;unsumed\;index}}\linebreak -m\frac{[X^i,X^j]^2}{l^4} +
O(X^6)\cdots,
\eea
in which $l$ is a parameter with dimension of length. Consequently, the
action (\ref{SX}) will be found to be the (low-energy bosonic) action of
$N$ D0-branes in 1-form RR field background $(A_t,A_i)$, in the ``temporal
gauge"  $a_0(t)=0$. From the String Theory point of view, D0-branes are
point particles to which ends of strings are attached \cite{9510017,
9611050}. In a bound state of $N$ D0-branes, D0-branes are connected to
each other by strings stretched between them, and it can be shown that the
correct dynamical variables describing the positions of D0-branes, rather
than numbers, are $N\times N$ hermitian matrices \cite{9510135}. By
restoring the (world-line) gauge potential $a_0(t)$, we conclude by the
action \cite{9910053,9910052}
\bea\label{SD0}
S_{{\rm D0}}= \int dt\; \tr \bigg(\haf m D_tX_i
D_t X^i + q D_tX^i A_i(X,t) - q A_t(X,t)+m\frac{[X^i,X^j]^2}{l^4}+
\cdots\bigg)
\eea
with $D_t=\partial_t+i[a_0(t),\;\;]$ as covariant derivative. 
Ignoring for the moment the gauge potentials $(A_t,A_i)$, the equations of
motion can be solved by diagonal configurations, such as:
\bea
X^i(t)&=&{\rm diag.} (x^i_1(t),\cdots, x^i_N(t)),\nonumber\\
a_0(t)&=&{\rm diag.} (a_{01}(t),\cdots, a_{0N}(t)),
\eea
with $x^i_\alpha(t)=x^i_{\alpha 0}+ v^i_\alpha t$, $\alpha=1,\cdots, N$. 
By this configuration, we restrict the $U(N)$ generators $T^a$ to the $N$
dimensional Cartan (diagonal) sub-algebra; saying with respect to symmetry
issues, the symmetry is broken from $U(N)$ to $U(1)^N$. This configuration
describes the classical free motion of $N$ D0-branes, neglecting the
effects of the strings stretched between them. Of course the situation is
different when we consider the quantum effects, and consequently it will
be found that the dynamics of the off-diagonal elements capture the
oscillations of the stretched strings. 

It can be seen that the transformations (\ref{UOT}), also leave 
the action (\ref{SD0}) invariant. By replacements one finds \cite{0103262}
\bea\label{VSD0}
\delta S_{{\rm D0}} \sim \delta S[X]+ q \int dt\; \tr 
\bigg(i a_0[X^i,\delta_i\Lambda(X,t)]\bigg)=0.
\eea
In above, $\delta S[X]$ is the expression introduced in (\ref{VSX}), and
the second term vanishes by the symmetrization prescription
\cite{0103262}.

{\bf Non-Abelian Transformations:} Actually, the action (\ref{SD0}) is
invariant under the transformations
\bea
X^i&\goes&\tilde X^i=U^\dagger(X,t)X^i U(X,t),\nonumber\\
a_0(t)&\goes& \tilde a_0(X,t)=U^\dagger(X,t) a_0(t) U(X,t) -i
U^\dagger(X,t)
\partial_t U(X,t),
\eea
with $U(X,t)$ as an arbitrary $N\times N$ unitary matrix;
in fact under these transformations one obtains
\bea\label{DTF1}
D_tX^i&\goes&\tilde
D_t\tilde X^i=U^\dagger(X,t)D_tX^iU(X,t),\\
\label{DTF2}
D_tD_tX^i&\goes&\tilde D_t \tilde D_t
\tilde X^i=U^\dagger(X,t)D_tD_tX^iU(X,t).
\eea
Now, in the same spirit as for the previously introduced $U(1)$ symmetry of
eq.(\ref{UOT}), one finds the symmetry transformations: 
\bea\label{NAT}
X^i&\goes&\tilde X^i=U^\dagger(X,t)X^i U(X,t),\nonumber\\
a_0(t)&\goes&\tilde a_0(X,t)=U^\dagger(X,t) a_0(t) U(X,t) -i
U^\dagger(X,t) \partial_t U(X,t),\nonumber\\
A_i(X,t)&\goes& \tilde A_i(X,t)=
U^\dagger(X,t)A_i(X,t)U(X,t)+iU^\dagger(X,t)
\delta_i U(X,t),\nonumber\\
A_t(X,t)&\goes& \tilde A_t(X,t)=
U^\dagger(X,t)A_t(X,t)U(X,t)-iU^\dagger(X,t)
\partial_t U(X,t),
\eea
in which we assume that $U(X,t)=\exp(-i\Lambda)$ is arbitrary up to this
condition that $\Lambda(X,t)$ is totally symmetrized in the $X$'s.  The
above transformations on the gauge potentials are similar to those of
non-Abelian gauge theories, and we mention that it is just the consequence
of enhancement of degrees of freedom from numbers ($x$) to matrices ($X$). 
In other words, we are faced with a situation in which ``the rotation of
fields" is generated by ``the rotation of coordinates." 

The above observation on gauge theory associated to D0-brane matrix
coordinates on its own is not a new one, and we already know another
example of this kind in non-commutative gauge theories. In spaces whose
coordinates satisfy the algebra
\bea
[\hat x^\mu,\hat x^\nu]=i\theta^{\mu\nu}, 
\eea 
with constant $\theta^{\mu\nu}$, the symmetry transformations of the
$U(1)$ gauge theory are like those of non-Abelian gauge theory
\cite{9908142,CDS,jabbari}, in the explicit form
\bea
A_\mu(x)\goes A'_\mu(x)=U^\dagger(x)\star A_\mu(x)\star U(x) 
-iU^\dagger(x)\star \partial_\mu U(x),
\eea
in which the $\star$-products are recognized. Also, one could put things
in the reverse direction that we had in above for D0-branes. The
coordinates
$x^\mu$ can be transformed locally by the large symmetry of the space as
$\tilde{x}^\mu \equiv U(x)\star x^\mu\star U^\dagger(x)$ \footnote{Here,
we are using $\hat x$ for operators as coordinates, and $x$ as numbers
multiplied by the $\star$-products.}. Note that the above
comutation relation is
satisfied also by the transformed coordinates. Now, by combining the
gauge transformations with a transformation of coordinates one can bring
the transformations of gauge fields to the form of a $U(1)$ theory, as
\bea
x^\mu &\goes&\tilde{x}^\mu= U(x)\star x^\mu\star
U^\dagger(x),\nonumber\\
A_\mu(x)&\goes& \tilde A_\mu(\tilde x)=A'(\tilde x)=A_\mu(x) 
-\partial_\mu \Lambda(x),
\eea
with $U=\exp(-i\Lambda)$. One also notes that by the above transformation
the so-called ``covariant coordinates" 
${\cal{X}}^\mu\equiv x^\mu +(\theta^{-1} A(x))^\mu$ remain invariant.
In addition, the case we see here for D0-branes
may be considered as another example of the relation between gauge
symmetry transformations and transformations of matrix coordinates
\cite{0007023}.

The last notable points are about the behaviour of $a_0(t)$ and $A_t(X,t)$
under symmetry transformations (\ref{NAT}). From the world-line theory
point of view, $a_0(t)$ is a dynamical variable, but $A_t(X,t)$ should be
treated as a part of background, however they behave similarly under
transformations. Also we see by (\ref{NAT}) that the time, and only time
dependence of $a_0(t)$, which is the consequence of dimensional reduction,
should be understood up to a gauge transformation. In \cite{0103262} a
possible map between the dynamics of D0-branes, and the semi-classical
dynamics of charged particles in Yang-Mills background was mentioned. It
is worth mentioning that via this possible relation, an explanation for
the above notable points can be recognized \cite{0103262}.

{\bf Lorentz Equations Of Motion:} The equations of motion by action
(\ref{SD0}), ignoring for the moment the potential term $V(X)$, will be
found to be
\bea\label{LORENZ}
&~&mD_tD_t X_i=q\bigg(E_i(X,t)
+\underbrace{D_tX^jB_{ji}(X,t)}\bigg),\\
\label{A0EOM}
&~&m[X_i,D_tX^i]=q[A_i(X,t),X^i],
\eea
with the following definitions
\bea
\label{ELEC}
E_i(X,t)&\equiv&-\delta_i A_t(X,t)-\partial_t A_i(X,t),\\
\label{MAGN}
B_{ji}(X,t)&\equiv&-\delta_jA_i(X,t)+\delta_iA_j(X,t).
\eea
In above, the symbol $\underbrace{D_tX^j B_{ji}(X,t)}$ denotes the average
over all of positions of $D_tX^j$ between the $X$'s of $B_{ji}(X,t)$. The
above equations for the $X$'s are like the Lorentz equations
of motion, with the exceptions that two sides are $N\times N$ matrices,
and the time derivatives $\partial_t$ are replaced by their covariant
counterpart $D_t$ \footnote{$D_t$ is absent in the definition of $E_i$,
because, the combination $i[a_0,A_i]$ has been absorbed to produce
$D_tX^j$ for both parts of $B_{ji}$.}.

The behaviour of eqs. (\ref{LORENZ}) and (\ref{A0EOM}) under gauge
transformation (\ref{NAT}) can be checked. Since the action is invariant
under (\ref{NAT}), it is expected that the equations of motion change
covariantly. The left-hand side of (\ref{LORENZ}) changes to $U^\dagger
D_tD_tX U$ by (\ref{DTF2}), and therefore we should find the same change
for the right-hand side. This is in fact the case, since
\bea
f(X,t) &\goes& \tilde{f}(\tilde{X},t)=
U^\dagger(X,t) f(X,t)U(X,t),\nonumber\\
\delta_i f(X,t)&\goes& 
\tilde{\delta}_i\tilde{f}(\tilde{X},t)=
U^\dagger(X,t) \delta_i f(X,t)U(X,t),\nonumber\\
\partial_t f(X,t)&\goes&
\partial_t \tilde{f}(\tilde{X},t)=
U^\dagger(X,t) \partial_t f(X,t) U(X,t).
\eea
In conclusion, the definitions (\ref{ELEC}) and
(\ref{MAGN}), lead to
\bea
E_i(X,t) &\goes& \tilde{E_i}(\tilde{X},t)=
U^\dagger(X,t) E_i(X,t)U(X,t),\nonumber\\
B_{ji}(X,t) &\goes& \tilde{B}_{ji}(\tilde{X},t)=
U^\dagger(X,t) B_{ji}(X,t)U(X,t),
\eea
a result consistent with the fact that $E_i$ and $B_{ji}$ are functionals
of $X$'s. We thus see that, in spite of the absence of the usual
commutator term $i[A_\mu,A_\nu]$ of non-Abelian gauge theories, in our
case the field strengths transform like non-Abelian ones. We recall that
these are all consequences of the matrix coordinates of D0-branes. 
Finally by the similar reason for vanishing the second term of
(\ref{VSD0}), both sides of (\ref{A0EOM}) transform identically.

An equation of motion similar to (\ref{LORENZ}) is considered in
\cite{fat241,fat021} as a part of similarities between the dynamics of
D0-branes and bound states of quarks--QCD strings
\cite{fat241,fat021,02414}. The point is that the center-of-mass dynamics
of D0-branes is not affected by the non-Abelian sector of the background,
\ie the center-of-mass is ``white" with respect to $SU(N)$ sector of
$U(N)$. The center-of-mass coordinates and momenta are defined by: 
\bea 
X^i_{c.m.}\equiv \frac{1}{N}
\tr X^i,\;\;\;\;  P^i_{c.m.}\equiv \tr P^i_X, 
\eea 
where we are using the convention $\tr {\bf 1}_N=N$. To specify the net
charge of a bound state, its dynamics should be studied in zero magnetic
and uniform electric fields, \ie$B_{ji}=0$ and $E_i(X,t)=E_{0i}$
\footnote{In a non-Abelian gauge theory an uniform electric field can be
defined up to a gauge transformation, which is quite well for
identification of white (singlet) states.};  thus these fields are not
involved by $X$ matrices, and contain just the $U(1)$ part. In other
words, under gauge transformations $E_{0i}$ and $B_{ji}=0$ transform to
$\tilde{E}_i(X,t)=V^\dagger(X,t)E_{0i}V(X,t)=E_{0i}$ and
$\tilde{B}_{ji}= 0$. Thus the action (\ref{SD0}) yields the following
equation of motion: 
\bea
(Nm_0){\ddot{X}}^i_{c.m.}=Nq E^i_{0(1)},
\eea
in which the subscript (1) emphasises the $U(1)$ electric field.  So the
center-of-mass only interacts with the $U(1)$ part of $U(N)$. From the
String Theory point of view, this observation is based on the simple fact
that the $SU(N)$ structure of D0-branes arises just from the internal
degrees of freedom inside the bound state. 

{\bf Map To Non-Abelian:} In \cite{9908142} a map between field
configurations of non-commutative and ordinary gauge theories is
introduced, which preserves the gauge equivalence relation. It is
emphasized that the map is not an isomorphism between the gauge groups. 
It will be interesting to study the properties of the map between
non-Abelian gauge theory and gauge theory associated with matrix
coordinates of D0-branes; on one side the quantum theory of matrix fields,
and on the other side the quantum mechanics of matrix coordinates. Since
in this case we have matrices on both sides, it may be possible to find an
isomorphism between all objects involving in the two theories, \ie
dynamical variables and transformation parameters.

It is useful to do some imaginations in this direction. We may begin by
the action 
\bea
S=\int d^{d+1}x \bigg(-\bar\psi (\gamma^\mu 
\partial_\mu +m)\psi-\tr(\frac{1}{4g^2} F_{\mu\nu}F^{\mu\nu}
+J_\mu A^\mu)\bigg),\\
A^\mu(x)=A^\mu_a(x)T^a,\;\;\;
F^{\mu\nu}(x)=F^{\mu\nu}_a(x)T^a,\;\;\mu,\nu=t,i,\nonumber
\eea
in which the term $J_\mu A^\mu$ is responsible for the interaction, and
can be taken the standard form $J_\mu^a=i\bar\psi\gamma_\mu T^a \psi$.
Gauge invariance specifies the behaviour of the current $J_\mu$ under the
gauge transformations to be $J(x)\goes J'(x)=U^\dagger J(x) U$. 

Now we can sketch the form of the map between two theories as follows
\vspace{.3cm}
\begin{center}
\begin{tabular}{c c c}
Non-Abelian Gauge Theory & {\large $\Leftrightarrow$} & 
Electrodynamics On Matrix Space\\\hline
$A^\mu(x)=A^\mu_a(x)T^a$ & {\large =} & $A^\mu(X) + 
(\Lambda(X)+\delta\Lambda(X))$\\
$F^{\mu\nu}(x)=F^{\mu\nu}_a(x)T^a$ & {\large =} &
$F^{\mu\nu}(X)$\\
$J^\mu(x)=J^\mu_a(x) T^a$ &  {\large =} & $D_\tau X^\mu$\\
$\Lambda(x)=\Lambda_a(x)T^a$ &{\large =} & $\Lambda(X)$
\end{tabular}
\end{center}
\vspace{.3cm} 
Two points should be emphasized. First, in above we are sketching the
relation or map between a field theory and a world-line theory of a
particle in a matrix space; like the same that we assume for relation
between field theories and theories living on the world-sheet of strings. 
Second, though other gauges (like the light-cone one of
\cite{fat021,fat241})
maybe have some more advantages, here we have assumed that a covariant
theory on matrix space is also available; in above it is needed to define
covariant derivative $D_\tau$ along the world-line (see \cite{0103262} as
an example of such a theory). In above table we mention that, firstly, the
objects in both sides are matrices, and so the number of degrees of
freedom matches.  Secondly, field strengths and currents of the two
theories transform identically, \ie in adjoint representation. 

The fate of the map after quantization is interesting. It remains to be
understood that which correlation functions of the two theories should be
put ``equal". We leave it for the future.

As the last point in this part, it will be interesting to mention the
conceptual relation between the above map, and the ideas concerned in
special relativity. Let us take the following general prescription in our
physical theories, that {\it the structure of space-time has to be in
correspondence with the fields}, saying: 
\begin{center}
{\large{\bf Fields $\Leftrightarrow$ Coordinates}}
\end{center}
In this way one understands that the space-time coordinates $x^\mu$ as
well as gauge potentials $A^\mu$ behave like a (d+1)-vector (spin 1) under
the boost transformations. This is just the same idea of special
relativity to change the picture of space-time such as to be consistent
with the Maxwell equations.

Also in this way supersymmetry is a natural continuation of the special
relativity program: Adding spin $\frac{1}{2}$ sector to the coordinates of
space-time, as the representatives of the fermions of nature. This leads
one to the super-space formulation of the supersymmetric theories, and in
the same way fermions are introduced into the bosonic String Theory. 

Now, what may be modified if nature has non-Abelian (non-commutative) 
gauge fields? In the present nature non-Abelian gauge fields can not make
spatially long coherent states; they are confined or too heavy. But the
picture may be changed inside those regions of space-time where such
fields are non-zero.  In fact recent developments of String Theory sound
this change and it is understood that non-commutative coordinates and
non-Abelian gauge fields are two sides of one coin. We may summarise the
above discussion in the table below \cite{fat021,fat241}. 

\vspace{.3cm}
\begin{center}
\begin{tabular}{|c|c|c|}
\hline Field & Space-Time Coordinates& Theory \\\hline
\hline Photon $A^\mu$ & $x^\mu$ & Electrodynamics  \\
\hline Fermion  $\psi$ &  $\theta$, $\bar\theta$ & Supersymmetric\\
\hline Gluon $A^\mu_a$ &  $X^\mu_a$ & Chromodymamics?\\
\hline
\end{tabular}
\end{center}
\vspace{.3cm} 

{\bf Finite $N$ Non-Commutative Phenomenology:} Recently non-commutative
field theories have attracted a large interest. Most of these kinds of
studies are concerning theories which are defined on spaces whose
coordinates satisfy the algebra: $[\hat x^\mu,\hat x^\nu]
=i\theta^{\mu\nu}$. This algebra is satisfied just by $\infty\times\infty$
matrices, and as the consequence, the concerned non-commutativities should
be assumed in all regions of the space. Also, generally in these spaces
one should expect violation of Lorentz invariance.

In the case we have for D0-branes, the non-commutativity of
matrix coordinates is ``confined" inside the bound state, and so it
appears to be different, and maybe more interesting. How can we probe this
non-commutativity? The answer is gained simply through ``the response of
non-commutativity to the external probes." The dynamics of D0-branes in
background of curved metric $G_{\mu\nu}(x,t)$ and the 1-form (RR) field
$A_\mu(x,t)$ can be given in lowest orders by (not being very precise
about indices and coefficients) \cite{9910053,9910052}:
\bea
S&=&\int dt \tr \bigg(\frac{m}{2} G_{ij}(X,t) D_tX^iD_t X^j 
+ q G_{ij}(X,t)A^i(X,t) D_tX^j- q A_t(X,t)
\nonumber\\
&~& +mG(X,t)G(X,t)\frac{[X,X]^2}{l^4}+ (1-G_{00}(X,t))+\cdots\bigg).
\eea
We again mention that the backgrounds $G_{\mu\nu}(x,t)$ and $A_\mu(x,t)$
appear in the action by functional dependence on matrix coordinates
$X$'s. In fact this is the key of ``How to probe non-commutativity?".
In a Fourier expansion of the background we find:
\bea
A(X,t)&=& \sum_k \bar A (k,t)\e^{ik_i X^i},\nonumber\\
G(X,t)&=& \sum_k \bar G (k,t)\e^{ik_i X^i},
\eea
in which $\bar A(k,t)$ and $\bar G(k,t)$ are the Fourier components of the
fields $A(x,t)$ and $G(x,t)$ respectively; \ie fields by ordinary
coordinates.  One can imagine the scattering processes which are designed
to probe inside the bound states. Such as every other scattering process
we have two regimes: 1) long wave-length, 2) short wave-length. 

In small $k$ (long wave-length) regime, the fields $A_\mu$ and
$G_{\mu\nu}$ are not involved by $X$ matrices mainly, and the fields will
appear to be nearly constant inside the bound state. So in this regime
non-commutativity will not be seen; Fig.1. 

\begin{figure}[h]
\begin{center}
\leavevmode
\epsfxsize=50mm
\epsfysize=30mm
\epsfbox{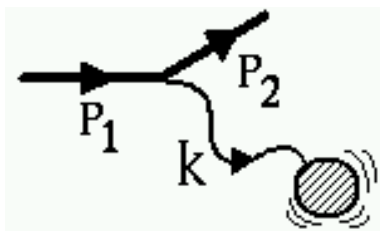}
\caption{Long wave-length scattering: sub-structure is not seen.}
\end{center}
\end{figure}

In the large $k$ (short wave-length) regime, the fields depend on
coordinates $X$, and so the sub-structure responsible for
non-commutativity should be probed; Fig.2. As we recalled previously, in
fact it is understood that the non-commutativity of D0-brane coordinates
is the consequence of the strings which are stretched between D0-branes.
So, by these kinds of scattering processes one should be able to probe
both D0-branes (as point-like objects), and the strings stretched between
them. 
\begin{figure}[h]
\begin{center}
\leavevmode
\epsfxsize=50mm
\epsfysize=30mm
\epsfbox{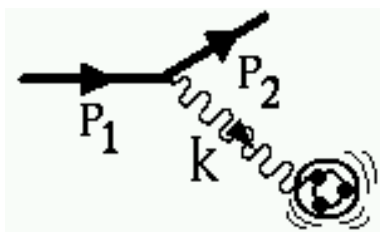}
\caption{Short wave-length scattering: non-commutativity is
probed.}
\end{center}
\end{figure}

{\bf Acknowledgment:} I am grateful to M. Hajirahimi for her careful
reading of the manuscript.


\end{document}